\documentclass[%
 reprint,
 nofootinbib,
superscriptaddress,
 amsmath,amssymb,
 aps,
]{revtex4-2}

\usepackage{graphicx}
\usepackage{dcolumn}
\usepackage{bm}
\usepackage[colorlinks=true,  urlcolor=blue,  linkcolor=blue,  citecolor=blue]{hyperref}

\usepackage{multirow}

\let\jnlstyle=\rmfamily 
\def\refjnl#1{{\jnlstyle#1}}%
\newcommand\pnas{\refjnl{PNAS}}
\newcommand\jmlr{\refjnl{JMLR}}
\newcommand\aj{\refjnl{AJ}}
\renewcommand\apj{\refjnl{ApJ}}
\newcommand\apjl{\refjnl{ApJL}}     
\newcommand\aap{\refjnl{A\&A}}
\newcommand\mnras{\refjnl{MNRAS}}
\renewcommand\prd{\refjnl{PhRvD}}

\newcommand\jcap{\refjnl{JCAP}}

\usepackage{tikz}
\usetikzlibrary{positioning, shapes.geometric}

\newcommand{\simbig}{{\textsc{SimBIG}}}

\begin{document}

\title{Mitigating Model Misspecification in Simulation-Based Inference \\ for Galaxy Clustering}
\author{Sébastien Pierre}
 \email{sebastien.pierre@phys.ens.fr}
\affiliation{Laboratoire de Physique de l’École Normale Supérieure, ENS, Université PSL,
75005 Paris, France}
\affiliation{Center for Computational Mathematics, Flatiron Institute, 162 5\textsuperscript{th} Avenue, New York, NY 10010, USA}

\author{Bruno Régaldo-Saint Blancard}
\email{bregaldo@flatironinstitute.org}
\affiliation{Center for Computational Mathematics, Flatiron Institute, 162 5\textsuperscript{th} Avenue, New York, NY 10010, USA}

\author{ChangHoon Hahn}
\affiliation{Steward Observatory, University of Arizona, 933 N. Cherry Avenue, Tucson, AZ 85721, USA}
\affiliation{Department of Astronomy, The University of Texas at Austin, 2515 Speedway, Stop C1400, Austin, TX 78712, USA}

\author{Michael Eickenberg}
\affiliation{Center for Computational Mathematics, Flatiron Institute, 162 5\textsuperscript{th} Avenue, New York, NY 10010, USA}

\date{\today}

\begin{abstract}
Simulation-based inference (SBI) has become an important tool in cosmology for extracting additional information from observational data using simulations.
However, all cosmological simulations are approximations of the actual universe, and SBI methods can be sensitive to model misspecification --- particularly when the observational data lie outside the support of the training distribution.
We present a method to improve the robustness of cosmological analyses under such conditions.
Our approach first identifies and discards components of the summary statistics that exhibit inconsistency across related simulators, then learns a transformation that brings the observation back within the support of the training distribution.
We apply our method in the context of a recent {\sc SimBIG} SBI galaxy clustering analysis using the wavelet scattering transform (WST) summary statistic.
The original analysis struggled to produce robust constraints for certain subsets of WST coefficients, where the observational data appeared out-of-distribution (OOD) relative to the training data.
We show that our method enables robust cosmological inference and resolves OOD issues, while preserving most of the constraining power.
In particular, the improved {\sc SimBIG} WST analysis yields $\Lambda$CDM constraints of $\Omega_m = 0.32^{+0.02}_{-0.02}$ and $\sigma_8 = 0.80^{+0.02}_{-0.02}$, which are respectively $1.4\times$ and $3.1\times$ tighter than those from a standard perturbation-theory-based power spectrum analysis, confirming the significant information gain of WST summary statistics.
The proposed method is easily applicable to other cosmological SBI contexts and represents a step toward more robust SBI pipelines.
\end{abstract}

\maketitle


\section{\label{sec:level1}Introduction}

Simulation-based inference (SBI) has recently signaled a
major shift for cosmological analyses of large-scale structure (LSS)~\citep{Alsing2019, Cranmer2020}. Traditionally, LSS analyses have relied on Bayesian inference using perturbation-theory-based models and explicit analytic likelihoods~\cite[e.g.,][]{Ivanov2020}, typically assumed to be Gaussian. SBI offers an alternative approach: forward modeled simulations combined with flexible neural networks to replace explicit likelihoods. In doing so, it enables the extraction of additional non-linear and higher-order cosmological information, inaccessible with standard analyses. A major requirement of SBI is a forward model capable of producing synthetic data that faithfully mimic real observations. However, this modeling step inevitably involves approximations, which can introduce model misspecification. 
This paper proposes a method to mitigate
the impact of this model misspecification, in the context of SBI analyses using galaxy clustering.

Galaxy clustering encodes key cosmological information about the expansion history of the Universe and the growth of structure~\citep{sargent1977, kaiser1987, eisenstein1998, hamilton1998}, enabling us to probe the nature of dark energy and test theories of gravity. In particular, in the standard $\Lambda$CDM paradigm, galaxy
clustering precisely constrains $\Omega_m$ and $\sigma_8$, cosmological parameters that quantify the Universe’s matter density and the amplitude of matter density fluctuations, respectively. SBI analyses of galaxy clustering significantly increase the precision of $\Omega_m$ and $\sigma_8$ constraints by leveraging high-fidelity simulations to accurately model higher-order clustering down to non-linear scales~\citep[e.g.,][]{Hahn2023_pnas, Hahn2023_nature}. While these simulations can extend to regimes where perturbation theory breaks down, their accuracy is still limited beyond $k > 0.5~h/\mathrm{Mpc}$. This is because simulating galaxy clustering requires modeling the gravitational evolution of matter across scales spanning many orders of magnitude, as well as the complex baryonic processes of galaxy formation. Furthermore, additional modeling is required to account for systematics in the observed dataset, such as survey geometry, selection biases, etc. Generating synthetic observations is, needless to say, a challenging task that inevitably involves multiple approximations --- likely to cause model misspecification.

This challenge is further compounded by the fact that leveraging more informative summary statistics inherently increases the risk of model misspecification. While such statistics can extract additional cosmological information and improve constraints, they also impose stronger requirements on the simulations, which have to accurately model the observational features that the statistics are sensitive to.
Thus, there is a trade-off between informativeness and robustness --- a balance that must be carefully managed in practical applications. We emphasize that while SBI approaches explicitly contend with model 
misspecification, it impacts and can bias {\em any} inference approach.

Coming up with an end-to-end simulation pipeline that connects cosmological parameters to synthetic galaxy catalogs was the goal of \textsc{SimBIG}, which introduced a simulator explicitly designed for SBI~\cite{Hahn2023_jcap}. This framework enabled constraining cosmological parameters by leveraging a variety of summary statistics, thereby exploiting the non-Gaussian information they capture. It led to analyses of the Sloan Digital Sky Survey III Baryon Oscillation Spectroscopic Survey (BOSS;~\citealt{Eisenstein2011, Dawson2013}) galaxy sample using multiple clustering statistics: the power spectrum~\cite{Hahn2023_pnas}, the bispectrum~\cite{Hahn2024}, the wavelet scattering transform (WST)~\cite{SimBIG.WST}, a convolutional neural network compression of the galaxy field~\cite{Lemos2023}, the skewed spectra~\cite{Hou2024}, and the marked power spectrum~\cite{Massara2024}. Most of these studies outperformed the power spectrum baseline~\cite{Hahn2023_nature}, partly meeting the expectations of earlier Fisher forecasts~\cite{Hahn2020, Hahn2021, Eickenberg2022, Valogiannis2022a, Massara2023, Hou2023}.

Among these analyses, the WST study~\cite{SimBIG.WST} revealed signs of model misspecification. On test simulations, the inferred posteriors --- particularly for $\sigma_8$ --- were found to be sensitive to small-scale modeling assumptions, an issue that was largely mitigated by applying a scale cut. However, when applied to the BOSS observation, substantial variation in the predicted cosmological parameters across equally well-trained neural network architectures suggested that the observation may lie out-of-distribution (OOD) relative to the training simulations, raising concerns about the robustness of the inference. Furthermore, while parallel analyses using alternative summary statistics did not explicitly find such issues, this may reflect the limitations in their robustness tests rather than the absence of model misspecification.

Other recent works have further highlighted the importance of sensitivity analysis in SBI for galaxy clustering~\citep[e.g.,][]{Modi2024, Jo2025, Montel2025, Bayer2025}.
In particular, \cite{Modi2024}, using similar forward models as \simbig, found that while different choices in gravity models do not significantly impact SBI constraints, choices in the halo finder and the galaxy model can significantly bias the cosmological constraints. 
These works underscore the need to assess potential biases introduced by model assumptions.

Using the WST-based {\sc SimBIG} analysis as a case study, this paper develops and tests a general method for mitigating model misspecification in cosmological SBI pipelines.

This paper is organized as follows. In Section~\ref{sec:context}, we describe the context of our work, defining the type of model misspecification we aim to address and providing a brief overview of the {\sc SimBIG} framework, which enables SBI for galaxy clustering data. In Section~\ref{sec : Method}, we present our method for mitigating model misspecification. In Sect.~\ref{sec:results}, we present our results, and in Sect.~\ref{sec:conclusion}, we conclude this paper. Additional figures can be found in Appendix~\ref{app:add_figs}.

\section{Context}
\label{sec:context}

In this section, we formalize the notion of model misspecification in the context of SBI, with a focus on cosmological scenarios involving a single observational sample. We then ground this discussion in the practical setting of galaxy clustering analyses using the \textsc{SimBIG} framework, and introduce two diagnostics for detecting model misspecification in this context.

\subsection{Model Misspecification Formalism}
\label{sub:Model misspecification formalism}

\subsubsection{Definition}

SBI relies on simulations to capture the relationship between model parameters and data. The simulator takes the form of a probabilistic model that generates data $x \in  \mathbb{R}^{n}$ given a parameter $\theta \in  \Theta \subset \mathbb{R}^{d}$. It is represented by a parametric family of distributions $\{ \mathbb{P}_{\theta}, \theta \in \Theta\}$.  In contrast, the true data originate from an unknown process denoted by $\mathbb{G}$. A crucial assumption in SBI is that the simulator can faithfully reproduce the observed data. In practice, however, this assumption is rarely satisfied exactly. We say model misspecification occurs when:
\begin{equation}\label{Hard definition of model misspecification}
    \forall \theta \in \Theta, ~\mathbb{P}_{\theta} \neq \mathbb{G},
\end{equation}
meaning that no distribution in the simulator family exactly matches the true data-generating process. This formal definition is of course quite stringent as, in realistic settings, simulators are always misspecified to some degree. Moreover, this definition is impractical in most situations, as we typically have access only to samples from $\mathbb{P}_{\theta}$ and $\mathbb{G}$, and can therefore evaluate compatibility only in a statistical sense.

\subsubsection{Case of a Single Observation}

Detecting model misspecification in cosmology applications is particularly challenging due to the limited number of observations. In galaxy clustering analyses, for example, we typically have access to only a single observation --- one sample $x$ drawn from the true data-generating process $\mathbb{G}$. In this paper, we choose to focus on scenarios where model misspecification can be confidently detected even from a single sample. Specifically, we consider a sufficient condition for misspecification: when the observed data point lies outside the support of the simulator’s output distribution~\footnote{This distribution can be confused with the model evidence when marginalizing over parameters drawn from a prior distribution $p(\theta)$.} (or lies in a region of negligible probability under all $\mathbb{P}_\theta$), that is:

\begin{equation}
\exists\, x \sim \mathbb{G}~\textrm{such that}~\forall \theta \in \Theta, x \notin \mathrm{supp} (\mathbb{P}_{\theta}).\footnote{In this work, as in most practical situations, the support of $\mathbb{P}_\theta$ is approximated using a finite set of simulated samples.}
\end{equation}

\begin{figure}
\includegraphics[scale=0.6]{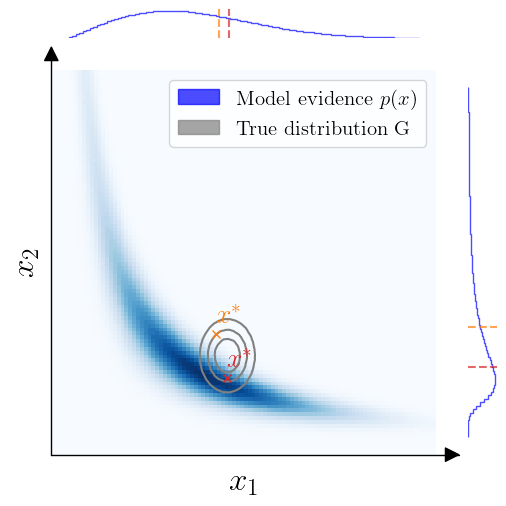}
\caption{Illustration of model misspecification in a 2D example. The simulator’s output distribution is shown in blue, and the true data-generating process $\mathbb{G}$, is represented in grey, revealing some misspecification. If the observed data corresponds to the red point --- lying within the simulator's support --- misspecification cannot be detected. In contrast, if the observed data is the orange point, which lies outside the simulator’s support, model misspecification can be confidently concluded. Interestingly, if we considered only the projection of the data onto $x_1$ or $x_2$ (a valid \textit{summary statistic}), both observations would fall within the marginal support, and misspecification would go undetected.}
\label{fig:model misspecification with one observation}
\end{figure}

Fig~\ref{fig:model misspecification with one observation} illustrates this idea with a simplified example. The blue region denotes the support of the simulator --- that is, the region of data space covered by model-generated samples --- while the grey contours represent the true data-generating process $\mathbb{G}$. Although the simulator is clearly misspecified relative to $\mathbb{G}$ in this example, this is not detectable if the only observed sample (red point) lies within the simulator’s support. Misspecification can only be confidently concluded when the observation (orange point) falls outside this support. This corresponds to an OOD scenario with respect to the simulation space.

\subsubsection{Summary Statistics}

Since the full simulated data $x \in \mathbb{R}^n$ is often high-dimensional, and the number of available simulations is typically limited by computational cost, it is common to map the data to a lower-dimensional space using a summary statistic. Formally, a summary statistic is a function $\nu: \mathbb{R}^n \rightarrow \mathbb{S}$, where $\mathbb{S}$ is a lower-dimensional space. The choice of summary statistic is critical, especially when dealing with potential OOD observations. As illustrated in Fig.~\ref{fig:model misspecification with one observation}, the detectability of misspecification can depend entirely on the choice of summary statistic. For the same data-generating process $\mathbb{G}$, an observation $x_{\text{obs}}$ may fall outside the support of the simulation set $\{x_i\}_{i=1}^{N_{\text{sim}}}$, where $N_{\text{sim}}$ is the number of simulations. However, its transformed version $\nu(x_{\text{obs}})$ might still lie within the support of the projected simulation set $\{\nu(x_i)\}_{i=1}^{N_{\text{sim}}}$. In this case, misspecification would go undetected. This highlights that different summary statistics may obscure or reveal model misspecification depending on how they map the observation relative to the simulated training distribution. In general, the more discriminative a summary statistic is --- i.e., the better it separates simulations from observations --- the more likely it is to reveal misspecification. At the same time, richer (e.g., higher-order) statistics can capture more cosmological information and lead to tighter parameter constraints. This introduces a trade-off between the \textit{informativeness} and the \textit{robustness} of the summary statistic. For the rest of this paper, we denote the summary statistic of a data point $x$ by $s = \nu(x)$.

\subsection{SBI from Galaxy Clustering Data with {\sc SimBIG}}
\label{sub:simbig}

This paper focuses on addressing model misspecification arising in the context of galaxy clustering SBI analyses. We work with the {\sc SimBIG} framework~\cite{Hahn2023_pnas}, which is tailored for analyzing a galaxy sample from SDSS-III BOSS. We refer to \cite{Hahn2023_pnas} for additional details on the target observational data.

\subsubsection{Forward Model}

The {\sc SimBIG} forward model maps cosmological parameters $\theta = (\Omega_{m},\Omega_{b},h,n_{s},\sigma_{8})$ to synthetic galaxy catalogs $x$~\cite{Hahn2023_jcap} using the following components: 
\begin{itemize}
    \item High-resolution \textsc{Quijote} $N$-body simulations, which yield the distribution of dark matter particles for different values of $\theta$.
    \item The dark matter halo finder \textsc{Rockstar}, which identifies dark matter halos from the simulation outputs.
    \item A halo occupation distribution (HOD) model, which populates halos with galaxies. This introduces 9 additional parameters, $\theta_{\rm HOD}$.
    \item Survey realism modeling, which applies the geometry, masking, and fiber collision effects of the BOSS galaxy sample.
\end{itemize}
The resulting synthetic data take the form of 3D point clouds, typically consisting of over 100,000 galaxies.

To make them suitable for inference, we compress the data into lower-dimensional summary statistics. Among the various summary statistics used in previous {\sc SimBIG} analyses, we choose to focus on the WST~\citep{SimBIG.WST}, for which model misspecification issues were particularly pronounced.
The analysis uses a total of 3,677 WST coefficients specifically tailored for galaxy surveys. They provide a multi-scale representation of the data by quantifying signal amplitudes at different oriented scales and capturing interactions between pairs of oriented scales. We refer the reader to \cite{SimBIG.WST} for the formal definition and detailed properties of these coefficients.

\subsubsection{Neural Posterior Estimation}

To perform inference on the cosmological parameters $\theta$ given summary statistics $s$, {\sc SimBIG} employs neural posterior estimation (NPE), a family of SBI techniques that approximate the posterior distribution $p(\theta \mid s)$ with neural density estimators, typically normalizing flows~\citep{Papamakarios2022, Kobyzev2021}.

In our setting, each simulation yields a pair $(\theta_i, s_i)$, where $\theta_i$ is drawn from the prior $p(\theta)$ and $s_i$ is the summary statistic extracted from a synthetic galaxy catalog generated by the forward model. A conditional neural density estimator $q_\phi(\theta \mid s)$, parameterized by $\phi$, is trained by minimizing the negative log-likelihood:
\begin{equation} \label{eq:loss_npe}
\mathcal{L}_{1}(\phi) = -\sum_{i} \log q_{\phi}(\theta_i \mid s_i).
\end{equation}

In practice, we use masked autoregressive flows (MAFs) as our density estimators --- a popular type of normalizing flow previously shown to be effective in this context. We train them using 20,000 synthetic galaxy catalogs generated with the {\sc SimBIG} forward model. Moreover, optimal MAF architectures are identified through a hyperparameter search, and the final estimator is built as a stacked model of the 10 best-performing architectures (based on validation loss)~\citep{Yao2023}. We refer to \cite{SimBIG.WST} for additional details on the training procedure.

\subsubsection{Mock Challenge}
\label{subsub: Mock Challenge}

In addition to the forward model, the {\sc SimBIG} framework includes a set of three test datasets designed to assess the robustness of the inference pipeline to alternative modeling choices. This setup constitutes the {\sc SimBIG} \textit{mock challenge}~\citep{Hahn2023_jcap}, which is comprised of:
\begin{itemize}
    \item \texttt{TEST0}: 500 galaxy catalogs generated using the original {\sc SimBIG} forward model. These catalogs are constructed at a fiducial cosmology, with $\theta_{\rm HOD}$ values sampled from a narrow prior.
    \item \texttt{TEST1}:  500 catalogs generated using an alternative forward model. These use the same $N$-body simulations as \texttt{TEST0}, but apply a different halo finder (FoF) and a simplified HOD model.
    \item \texttt{TEST2}: 1,000 catalogs generated using an alternative forward model based on \textsc{AbacusSummit} $N$-body simulations and the \textsc{CompaSO} halo finder. These simulations are constructed at a fiducial cosmology close to that of \texttt{TEST0}.
\end{itemize}
Since \texttt{TEST0} is generated by the same forward model as the training data, it allows us to test the pipeline on new, unseen data, where the model is correctly specified. In contrast, \texttt{TEST1} and \texttt{TEST2} are constructed using different simulation pipelines and thus serve as test cases for model misspecification. Additional details on these test sets can be found in~\citep{Hahn2023_jcap}.

\subsubsection{Diagnosing Model Misspecification}
\label{sec:diagnostics}

In the WST analysis conducted in \cite{SimBIG.WST}, two signs of model misspecification were observed: (1) biased parameter estimates when evaluated on simulations from \texttt{TEST1} and \texttt{TEST2}, and (2) high variability in the observational posteriors across neural density estimators with similar validation losses. These issues raised concerns about the robustness of the SBI pipeline as well as the reliability of the posterior distribution obtained from the observational data. We introduce two diagnostics designed to detect and quantify these issues.

\begin{figure}
\includegraphics[scale=0.3]{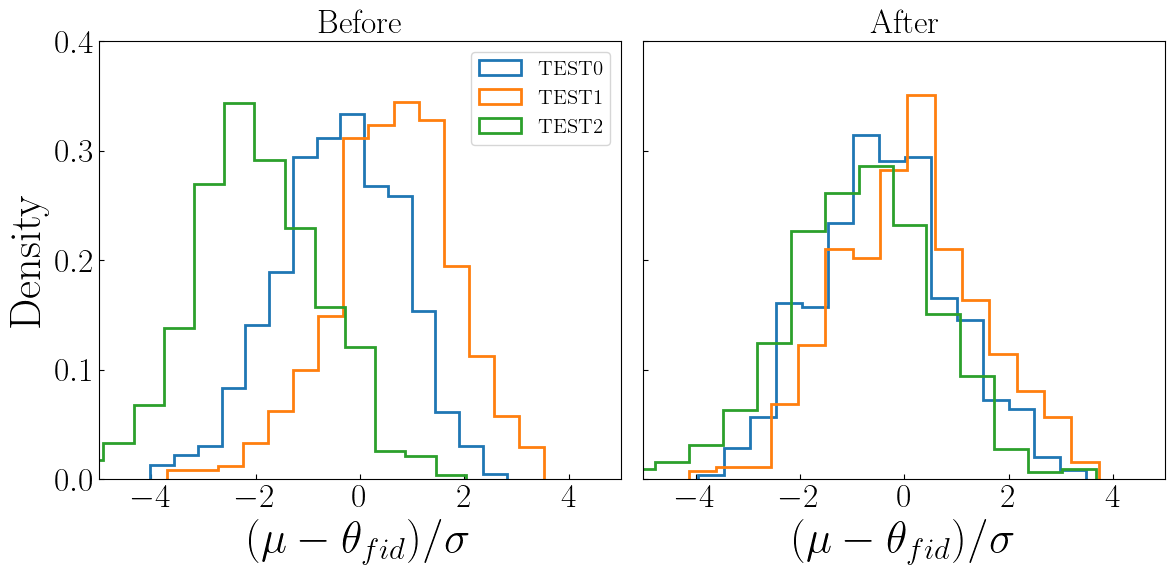}
\caption{Histograms of the normalized differences between the posterior mean $\mu$ and the true parameter value $\theta_{\mathrm{fid}}$ for the $\sigma_{8}$ parameter, computed as $(\mu - \theta_{\mathrm{fid}})/\sigma$. \texttt{Left}: Results before applying the method described in Sect.~\ref{sec : Method}, showing a lack of robustness in $\sigma_{8}$ inference across different  forward models. (Reproduced from \cite[][Fig.~6]{SimBIG.WST}.) \texttt{Right}: Results after applying our method. The three histograms are consistent, which indicates that the inference is now robust for simulations from both \texttt{TEST1} and \texttt{TEST2}.}
\label{fig:non robustness sigma8} 
\end{figure}

\paragraph{Robustness to the change of forward model}
\label{subsub: Robutness to the change of forward model}

To evaluate the robustness of our SBI pipeline to changes in the forward model, we compute the posterior mean $\mu$ and standard deviation $\sigma$ for each simulation of the {\sc SimBIG} mock challenge. We then examine the distribution of normalized deviations $(\mu - \theta_{\rm fid}) / \sigma$, where $\theta_{\rm fid}$ is the true cosmological parameter used to generate each simulation. We consider the pipeline to be robust if this distribution remains consistent across the different test datasets.

We reproduce in Fig~\ref{fig:non robustness sigma8} (left) the issue reported in \cite{SimBIG.WST}, which shows the distribution of normalized deviations for $\sigma_8$ before applying our method (introduced later in Section~\ref{sec : Method}). The figure shows that simulations from \texttt{TEST1} and \texttt{TEST2} yield biased $\sigma_8$ predictions when the pipeline is trained on simulations from the {\sc SimBIG} forward model, indicating a lack of robustness to changes in the simulator. Other parameters are not shown, as they do not exhibit significant robustness issues. This lack of robustness is concerning, as the actual observation may originate from a generative process that differs even more from the one used in training. If the inference is unstable under small changes to the forward model, our confidence in the resulting cosmological constraints is undermined. Passing this robustness test is thus a necessary condition for trustworthy inference.

\paragraph{Variability of posteriors}
\label{subsub: Variability of posteriors}

\begin{figure}
\includegraphics[scale=0.3]{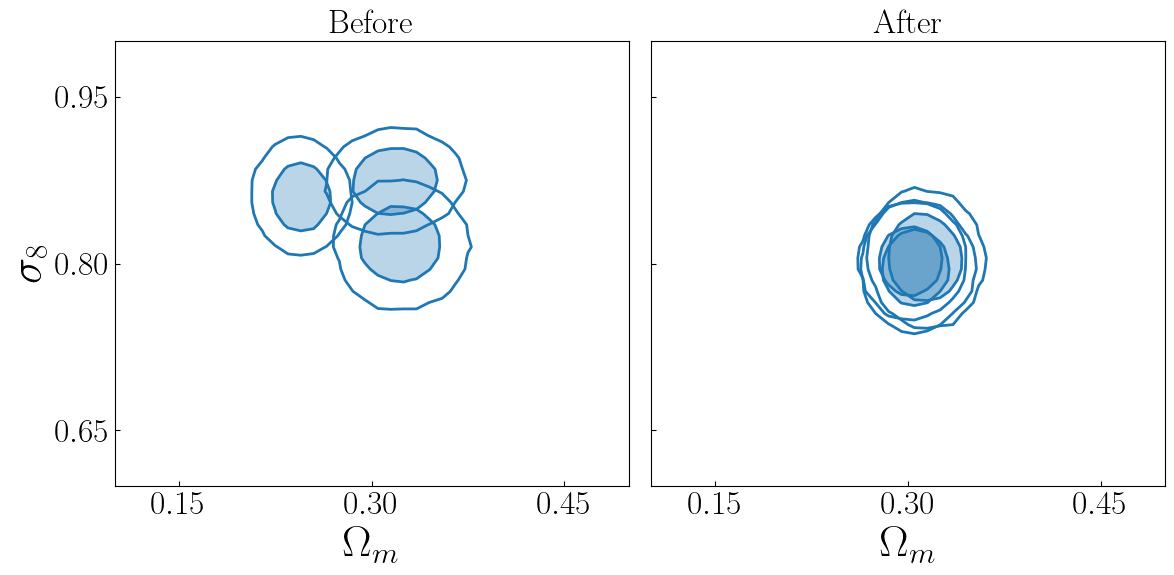}
\caption{Posterior predictions (68th and 95th percentile contours) for $(\sigma_{8}, \Omega_{m})$ obtained from the BOSS data using three different density estimator architectures, each achieving similar validation losses. \texttt{Left}: Before applying the method from Sect.~\ref{sec : Method}, the posteriors vary significantly across architectures for the same observation, suggesting that the data is OOD relative to the training set. (Results adapted from \cite[][]{SimBIG.WST}.) \texttt{Right}: After applying our method, the variability across architectures is substantially reduced, indicating that the observation is no longer OOD.}
\label{fig:variability sigma8 OmegaM} 
\end{figure}

While robustness to changes in the simulator is necessary, it is not sufficient: the observation itself may still be OOD with respect to the training data. We introduce a second diagnostic to detect this, based on the variability of inferred posteriors across neural density estimators. This test draws inspiration from the OOD detection literature \cite{Lakshminarayanan2016}. The intuition is that, for in-distribution data, different neural density estimators --- trained with the same data and achieving similar validation performance --- should yield similar posteriors from the observational data. In contrast, large variability across posteriors may indicate that the observation lies outside the training distribution.

In SBI, pipelines often involve a hyperparameter search over neural architectures, yielding multiple models with nearly equivalent validation losses. Despite their similar validation performance, these models may produce highly inconsistent posteriors when evaluated on OOD data. Fig~\ref{fig:variability sigma8 OmegaM} illustrates this effect using results from the {\sc SimBIG} WST analysis previously reported in \cite{SimBIG.WST}. It shows the 68th and 95th percentile contours of $(\sigma_8, \Omega_m)$ inferred by three neural density estimator architectures, each with comparable validation losses, applied to the BOSS galaxy sample. The substantial spread among the posteriors suggests that the observational data may be OOD relative to the training simulations. While the figure focuses on $(\sigma_8, \Omega_m)$ --- the most tightly constrained parameters --- other parameters also exhibit significant posterior variability, further supporting the presence of model misspecification.

To formalize this test, we define a variability score based on pairwise Kullback-Leibler divergences between posteriors. Let $\{q_{i}\}_{i=1}^{N{p}}$ denote the set of posteriors produced by $N_p$ different models. We define the variability score as:
\begin{equation}\label{eq:variability of posteriors}
    D_{v}(\{q_{i}\}_{i=1}^{N_{p}}) = \frac{1}{N_{p}(N_{p}-1)}\sum_{i\ne j} D_{\rm KL}(q_{i}|| q_{j}).
\end{equation}

The choice of divergence metric is not critical, provided it is easy to estimate; we verified that alternative standard choices yield similar trends. We consider the test to fail --- and thus signal potential model misspecification --- if the inferred posteriors are mutually incompatible, that is if $D_{v}$ is significantly larger than a relevant reference value. While passing the test does not guarantee well-specified models, failing it is a strong indicator of underlying issues. Alternatively, one can use the same architecture with different random initializations, as suggested in \cite{Lakshminarayanan2016}.\\

\begin{figure}
\includegraphics[scale=0.20]{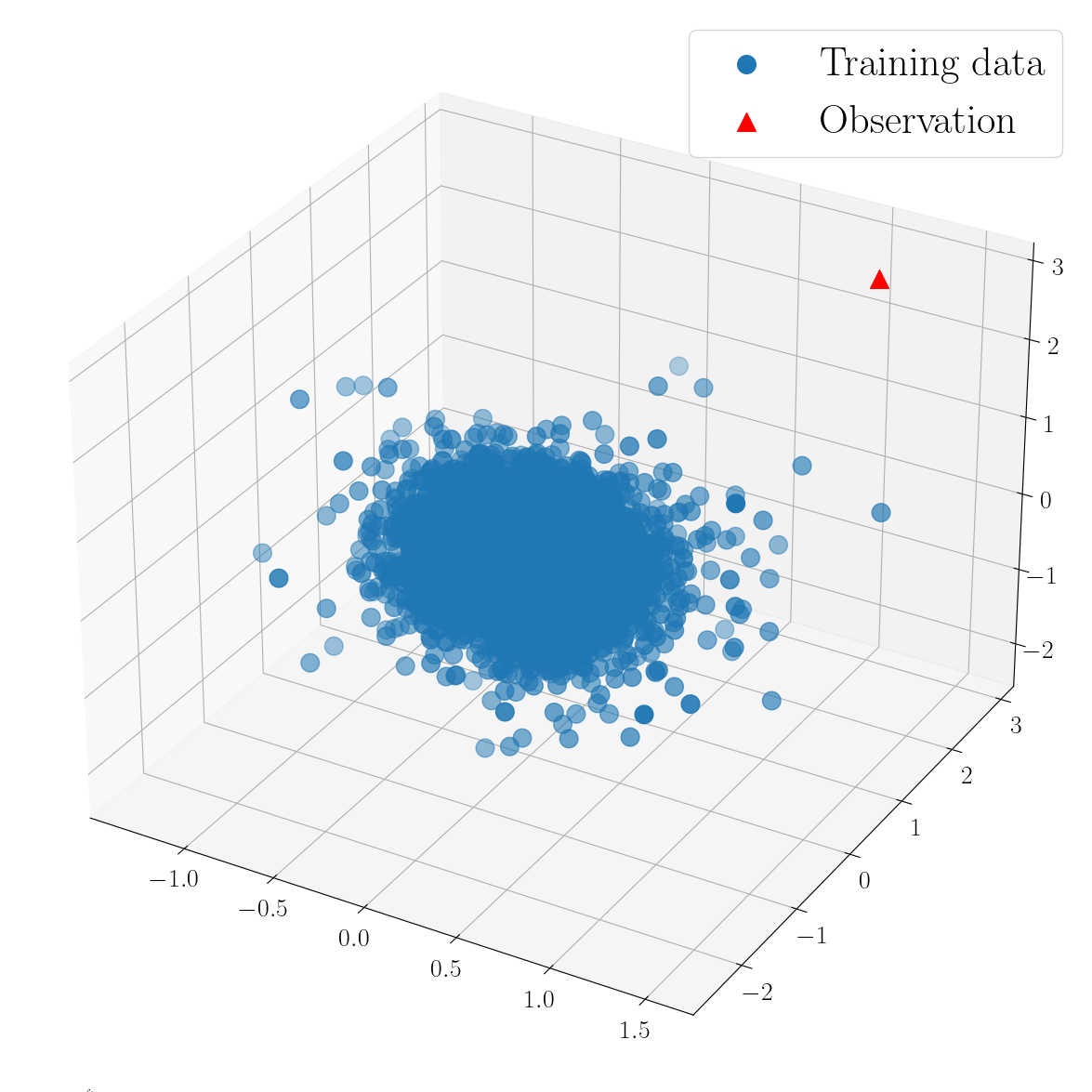}
\caption{3D linear projection of the training data (blue) and the observed data (red). The projection is obtained by performing PCA on the training set and projecting both the training and observed data onto the last three principal components. In this space, the observation is clearly OOD, indicating that it is also OOD in the full WST space.}
\label{fig:OOD_observation}
\end{figure}

Finally, we note that model misspecification can also be exhibited through standard dimensionality reduction techniques, although this alone does not clarify its impact on inference. For example, if the observational data appear OOD in a low-dimensional linear projection, they must also be OOD in the original high-dimensional space. Fig~\ref{fig:OOD_observation} illustrates this idea by showing the observational data (red point) projected onto the last 3 components of a PCA fitted on the training set (blue points). The observation clearly lies outside the distribution of the training data in this low-dimensional space, confirming its OOD nature in the original WST space.

\section{Method}
\label{sec : Method}

In this section, we introduce our method for addressing the model misspecification issues described in Section~\ref{sec:context}. The approach consists of a two-step procedure designed to discard information that causes the observation to lie outside the support of the training distribution. The first step is tailored to the {\sc SimBIG} framework and leverages the {\sc SimBIG} mock challenge datasets to apply a hard threshold on summary statistics that show inconsistency across closely related forward models. The second step is more general and applicable to any SBI pipeline. It is inspired by \cite{Robust_statistics_misspecification} and involves learning a transformation of the summary statistics to reduce OOD effects, drawing on ideas from the domain adaptation literature~\cite{Tzeng2014}.

\subsection{Discarding Simulator-Inconsistent Summary Statistics}
\label{sub : Cutting most misspecified information}

We make use of the test sets \texttt{TEST0} and \texttt{TEST1} (see Sect.~\ref{subsub: Mock Challenge}), which are generated from different simulators but share the same fiducial cosmological parameters. For each WST coefficient, we compare its marginal distribution across the two test sets. Significant discrepancies in these marginals indicate simulator disagreement, suggesting that the corresponding coefficient may encode simulator-specific artifacts rather than robust cosmological information. We exclude such coefficients from inference to reduce the impact of misspecification and improve robustness to modeling choices in the simulator.

\begin{figure}
\includegraphics[scale=0.6]{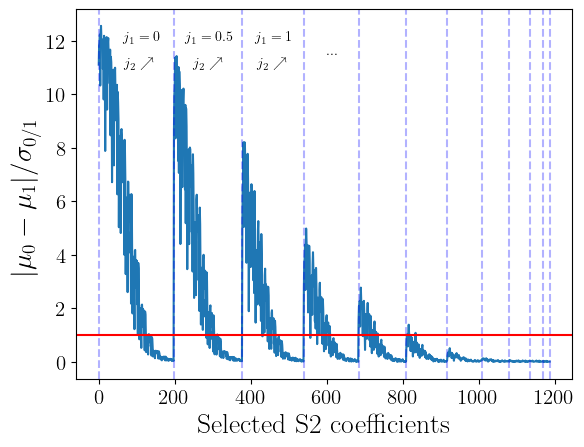}
\caption{Normalized mean differences between the marginal distributions of \texttt{TEST0} and \texttt{TEST1} for a subset of $S_2$ WST coefficients. These coefficients probe interactions between scales $j_1$ and $j_2$. Dashed lines indicate fixed values of $j_1$, ordered from smallest to largest scale. Within each segment, $j_2$ varies from smallest to largest scale. Coefficients above the red threshold line are discarded, as they exhibit significant differences between the two distributions. Most of the discarded information comes from small-scale features, though some small-scale information is retained through coefficients that capture interactions between small and larger scales.}
\label{fig:Difference statistics TEST0/TEST1} 
\end{figure}

In practice, we have 3,677  WST coefficients. For each individual coefficient, we compute the means and standard deviations of the \texttt{TEST0} and \texttt{TEST1} marginal distributions, denoted $\mu_{0}$, $\sigma_0$ and $\mu_{1}$, $\sigma_1$, respectively. Empirically, we find that $\sigma_0$ and $\sigma_1$ are nearly identical. We then compute the standardized difference for each coefficient as $\frac{|\mu_{0}-\mu_{1}|}{\sigma_{0/1}}$. Results for a representative subset of coefficients are shown in Fig.~\ref{fig:Difference statistics TEST0/TEST1}. For some coefficients, this normalized distance reaches up to 12$\sigma$, indicating a strong sensitivity to the choice of simulator. To mitigate this, we exclude any coefficient whose mean difference exceeds $1\sigma$, which removes just under half of the coefficients. Most of the discarded coefficients correspond to small-scale features --- a strategy previously shown to improve robustness in \cite{SimBIG.WST} --- although, in contrast to that work, not all small-scale coefficients are removed. This filtering step significantly improves robustness across simulators, including for \texttt{TEST2}. However, it is not sufficient on its own to pass the posterior variability test on the observation. This suggests that, while the method reduces simulator sensitivity, the observation itself may still be OOD.

\subsection{Robust Transformations of Summary Statistics}
\label{sub:robust transformation summary statistics}

The second step of our procedure improves robustness to model misspecification by learning a transformation of the summary statistics obtained in the first step, such that the observation is no longer OOD. This method is adapted from \cite{Robust_statistics_misspecification}, which regularizes the training of NPEs by penalizing transformations that lead to large discrepancies between the training data and the observation.

This learning-based approach is preferred over standard dimensionality reduction techniques (e.g., PCA), as it provides greater flexibility to explore the trade-off between robustness and informativeness. In particular, it enables non-linear transformations that can preserve more relevant information about the parameters than fixed linear projections.

\subsubsection{Principle}

We briefly summarize the approach and refer the reader to \cite{Robust_statistics_misspecification} for full details. Building on the NPE framework introduced in Section~\ref{sub:simbig}, we assume access to pairs $(\theta_i, s_i)$ sampled from the joint distribution $p(\theta, s)$. Standard NPE methods train a neural density estimator $q_\phi(\theta \mid s)$ to minimize the negative log-likelihood, as in Eq.~\eqref{eq:loss_npe}. However, when the observed summary $s_{\rm obs}$ lies far from the support of ${s_i}$, the resulting inference may be unreliable \cite{Cannon2022, SimBIG.WST}. To mitigate this, we learn a transformation of the summary statistics that reduces the distance between $s_{\rm obs}$ and the training data. This transformation, denoted $\eta :\mathbb{S} \rightarrow \mathbb{S'}$, can be selected from a parametric family $\{\eta_{\psi}\}$. Note that the transformed space $\mathbb{S'}$ may differ from $\mathbb{S}$, potentially being of lower dimensionality to explicitly discard misspecified or uninformative features.

Let $d$ be a distance between probability distributions. The objective is to minimize $d(\{\eta_{\psi}(s_{i})\}_{i=1}^{N},\{\eta_{\psi}(s_{\rm obs})\})$. However, minimizing this distance alone can lead to a degenerate solution in which $\eta_{\psi}$ collapses to a constant function, resulting in posteriors that coincide with the prior distribution. Therefore, a trade-off must be maintained: the transformation $\eta_{\psi}$ should both preserve information relevant for inferring $\theta$ and reduce the distance between the training and observed statistics. To achieve this balance, \cite{Robust_statistics_misspecification} proposes augmenting the standard NPE objective (Eq.~\eqref{eq:loss_npe}) with a regularization term based on the distance between the transformed training and observed statistics. This yields the modified loss:

\begin{align}\label{loss with regularization}
        \mathcal{L}_{2}(\phi,\psi) = & -\sum_{i} \log q_{\phi}(\theta_{i}|\eta_{\psi}(s_{i})) \\
        & + \lambda \, d(\{\eta_{\psi}(s_{i})\}_{i},\{\eta_{\psi}(s_{\rm obs})\}).        
\end{align}

Here, the parameter $\lambda > 0$ controls the strength of the regularization term and governs the trade-off between informativeness and robustness.

\subsubsection{Choice of Distance \texorpdfstring{$d$}{d}}

For practical implementation, we follow~\cite{Robust_statistics_misspecification} and use the squared maximum mean discrepancy (MMD) as our distance $d$. The MMD~\cite{Gretton12} is a kernel-based measure of the discrepancy between two probability distributions, $\mathbb{P}$ and $\mathbb{Q}$. It relies on a kernel function $k : \mathbb{S'} \times \mathbb{S'} \to \mathbb{R}$ that is symmetric and positive-definite. The squared MMD between $\mathbb{P}$ and $\mathbb{Q}$ is defined as:

\begin{align*} \mathrm{MMD}(\mathbb{P},\mathbb{Q})^{2}_{k} = &  \mathbb{E}_{s,s' \sim \mathbb{P}}[k(s,s')] + \mathbb{E}_{s,s' \sim \mathbb{Q}}[k(s,s')] \\
& - 2\mathbb{E}_{s \sim \mathbb{P}, s' \sim \mathbb{Q}}[k(s,s')].
\end{align*}

This quantity can be estimated even when the sample sizes of $\mathbb{P}$ and $\mathbb{Q}$ differ, which is critical in our case, where we compare a dataset to a single observation. Following \cite{Robust_statistics_misspecification}, we use a Gaussian kernel of the form $k(x,y) = \exp \left(-\lVert x-y \rVert ^{2} / 2l \right)$, where $l$ denotes the kernel's length scale. We determine $l$ using the median heuristic from~\cite{Gretton12}, setting it to the median of all squared pairwise distances within the training set. In our single-observation setting, $\mathbb{Q}$ reduces to a Dirac distribution, and the empirical squared MMD used during training simplifies to:

\begin{align*} \mathrm{MMD}^{2}(\{{\eta_{\psi}(s_{i})\}_{i}},& {\eta_{\psi}(s_{\rm obs})}) = \frac{1}{N^{2}}\sum_{i, j}e^{-\lVert \eta_{\psi}(s_{i})-\eta_{\psi}(s_{j})\rVert ^{2} / 2l}\\
& + 1 - \frac{2}{N}\sum_{i}e^{-\lVert \eta_{\psi}(s_{i})-\eta_{\psi}(s_{\rm obs})\rVert ^{2} / 2l}. 
\end{align*}

Note that due to the single-observation setting, minimizing this empirical MMD will not account for the full complexity of the true data-generating process $\mathbb{G}$. Without appropriate safeguards, this can lead to mapping the observation toward the centroid of the transformed training set, potentially introducing bias. In the next section, we present a strategy for selecting $\lambda$ that mitigates this issue.

\subsubsection{Setting \texorpdfstring{$\lambda$}{lambda}}
\label{subsub : Setting lambda}

\begin{figure}
\includegraphics[scale=0.5]{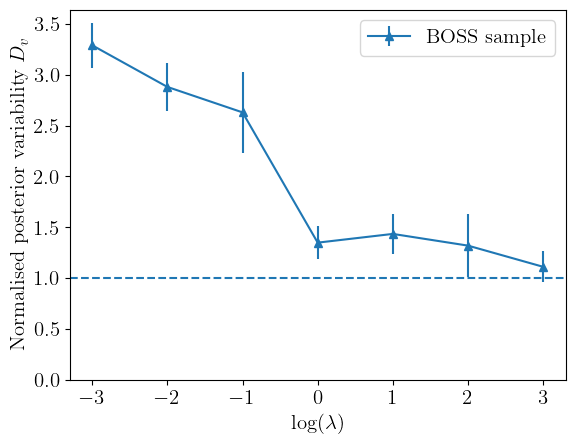}
\caption{Normalized posterior variability $D_{v}$ (see Sect.~\ref{sec:diagnostics}) across 10 different network architectures for the BOSS observation. The values are normalized by the mean posterior variability computed over a subset of the validation set. The normalized posterior variability drops sharply up to $\lambda = 1$, identifying it as the optimal value for our method. Beyond this point, it converges to 1.0, indicating that the posterior collapses to the prior for any input. Error bars are estimated via jackknife resampling.}
\label{fig:variability posterior lambda up}
\end{figure}

We now discuss the importance of selecting an appropriate value for $\lambda$ and the risks associated with suboptimal choices.

For $\lambda = 0$, loss of Eq.~\eqref{loss with regularization} reduces to the standard NPE objective of Eq.~\eqref{eq:loss_npe}, with no regularization. Conversely, setting $\lambda$ extremely high causes the learned summary statistics to become nearly constant across all data points, leading the posterior to collapse to the prior and rendering inference uninformative.

Selecting an appropriate $\lambda$ is therefore crucial. Fig~\ref{fig:variability posterior lambda up} shows how the variability metric $D_v$ (defined in Sect.~\ref{sec:diagnostics}) evolves with $\lambda$, measuring the spread of posteriors across multiple architectures for the observed data. For each $\lambda$ value, it is normalized by the mean posterior variability computed over a subset of the validation set. For the observed data, when $\lambda$ is very small, the variability of posteriors remains high, indicating that the posteriors are inconsistent (as in Fig.~\ref{fig:variability sigma8 OmegaM}, left). This suggests that the regularization is too weak, and the learned transformation $\eta$ fails to sufficiently adjust the summary statistics to bring the observation within the support of the training distribution. However, at a specific threshold of $\lambda$, we observe a clear drop in $D_{v}$, signaling that $\eta$ has effectively mapped the observation into the training distribution. Increasing $\lambda$ further shifts the posteriors toward the center of the prior. As $\lambda \to \infty$, the learned statistics degenerate, and posteriors collapse to the prior for all data points, leading to $D_{v} = 1$.

Based on this behavior, we select $\lambda$ as the smallest value at which $D_{v}$ drops significantly. This point corresponds to the minimal regularization needed to bring the observation in-distribution, while avoiding over-regularization that would bias the posterior toward the center of the prior. In Section \ref{subsub:Dependence on the prior}, we further demonstrate the robustness of this approach for different priors. \\

Finally, we emphasize the importance of applying both steps of our method to address the issues presented in Sect.~\ref{sec:diagnostics}. Indeed, applying only the regularization step described in Sect.~\ref{sub:robust transformation summary statistics} yields nearly identical posteriors for the observation but does not resolve robustness issues across different simulators. Conversely, using only the approach from~\ref{sub : Cutting most misspecified information} does not mitigate the high posterior variability across architectures, indicating that the observation remains OOD.

\section{Results}
\label{sec:results}

In this section, we present our results, including a validation of the method on synthetic data (Sect.~\ref{subsub:Validation on an observation coming from a different simulator}) and its application to the BOSS observation (Sect.~\ref{sec:obs_results}).

\subsection{Validation of the Method}
\label{subsub:Validation on an observation coming from a different simulator}

\subsubsection{Training Procedure}

We train 10 different MAF architectures using the hyperparameters identified as optimal in \cite{SimBIG.WST}.\footnote{Although our setting differs slightly --- due to the exclusion of some summary statistic coefficients and the addition of a new loss term --- we do not expect these changes to substantially alter the optimal hyperparameter configuration. Empirically, validation losses remain similar, justifying our choice to reuse the architectures and avoid the computational cost of a new hyperparameter search.} For $\eta_{\psi}$, we choose a multi-layer perceptron architecture, detailed in Fig.~\ref{fig:architecture}. Importantly, a key design goal was to prevent the transformation from being locally invertible around the observation, as otherwise, the $q_{\phi}$ conditioner network would have been able to learn to invert $\eta_{\psi}$ in the neighborhood of the observation, defeating the purpose of the transformation.\footnote{To verify non-invertibility of the final transformation, we inspected the singular values of the Jacobian of $\eta_{\psi}$ at the observation and found that approximately 60\% of them were zero.}

We regularize the optimization with an early stopping strategy focused on the first term of the loss of Eq.~\eqref{loss with regularization}. Specifically, training is halted once the validation loss stops improving over a fixed number of epochs. In practice, we note that accounting for the second term of the loss in this strategy does not significantly affect the overall training time. To set the regularization strength $\lambda$, we follow the procedure described in Section~\ref{subsub : Setting lambda} by exploring a range of values logarithmically spaced between $10^{-3}$ and $10^{3}$.

Finally, we define the final posterior density estimator as a stacked model formed by a linear mixture of the 10 trained MAFs:
\begin{equation}\label{mixture model}
q_{\rm ens}(\theta \mid s) = \sum_{i=1}^{10} w_{i} q_{\phi_{i}}(\theta \mid \eta_{\psi_{i}}(s)).
\end{equation}
The mixture weights ${w_i}$ are optimized to minimize the validation loss associated with this linear mixture, following~\cite{Yao2023}.

\subsubsection{Calibration}

\begin{figure}
\includegraphics[scale=0.4]{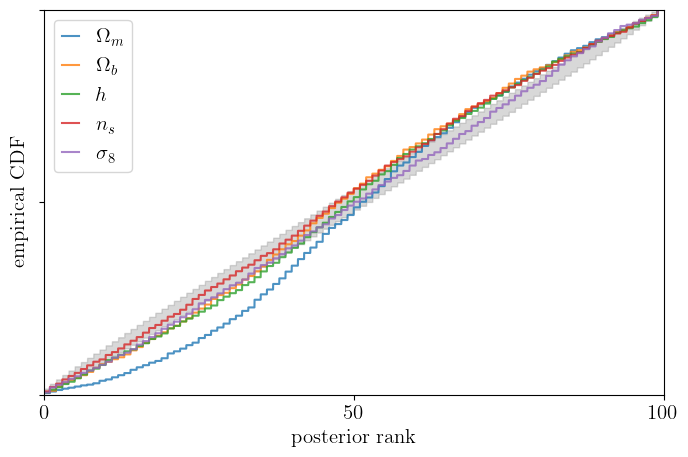}
\caption{Empirical cumulative distribution function of the SBC rank statistics for each cosmological parameter. For optimal calibration, the empirical cumulative distribution function of the rank statistics should lie within the grey region. Overall, our posteriors appear well calibrated, with an indication of slight underconfidence for $\Omega_{m}$.}
\label{fig:sbc} 
\end{figure}

As in previous {\sc SimBIG} analyses, we use simulation-based calibration (SBC) \cite{Talts2018} to assess the accuracy of our posterior estimates. Due to the limited data, we use the validation set in place of an independent test set. For each validation sample $(\tilde{\theta}, \tilde{s})$, we draw 100 posterior samples from $q_{\rm ens}(\theta \mid \tilde{s})$ and compute the rank of each component of $\tilde{\theta}$ among the corresponding posterior samples. Results for each cosmological parameter are shown in Fig.~\ref{fig:sbc}. For optimal calibration, the empirical cumulative distribution function of the rank statistics should lie within the grey region.

Overall, the posteriors appear well-calibrated. We observe mild deviations from the grey region for some parameters, particularly $\Omega_{m}$, where the posterior appears slightly underconfident. We attribute this to difficulties in training the models due to the limited size of the training dataset. While this underconfidence does not compromise the validity of our analysis, it suggests that the full constraining power of the WST may not yet be fully exploited.

\subsubsection{Robustness}

We assess the robustness of our inference method to changes in the simulator as described in Sect.~\ref{sec:diagnostics}. We note that while we indirectly made use of information from \texttt{TEST0} and \texttt{TEST1} through the hard cut on incompatible summary statistic coefficients, no information from \texttt{TEST2} was used during training or selection. Fig~\ref{fig:non robustness sigma8} (right) demonstrates the improved robustness of our pipeline, showing that the inferred results for $\sigma_8$ are now consistent across all test simulators, including \texttt{TEST2}. Although the figure focuses on $\sigma_8$, we observe similar robustness for the remaining cosmological parameters. This improvement is primarily driven by the hard coefficient selection described in Sect.~\ref{sub : Cutting most misspecified information}. However, without the learned transformation of the summary statistics, residual bias would remain for \texttt{TEST2}, highlighting the importance of both steps in the method.

\subsubsection{Application on Test Data}

We illustrate the impact of our method using synthetic data that mimics an observational setting where model misspecification must be addressed. Specifically, we randomly select a simulation from \texttt{TEST2}. As discussed in Section~\ref{sec:diagnostics}, inferences from \texttt{TEST2} without correction exhibit systematic bias --- particularly in $\sigma_8$ --- and show significant posterior variability, as measured by the $D_v$ metric.

\begin{figure}
\includegraphics[scale=0.28]{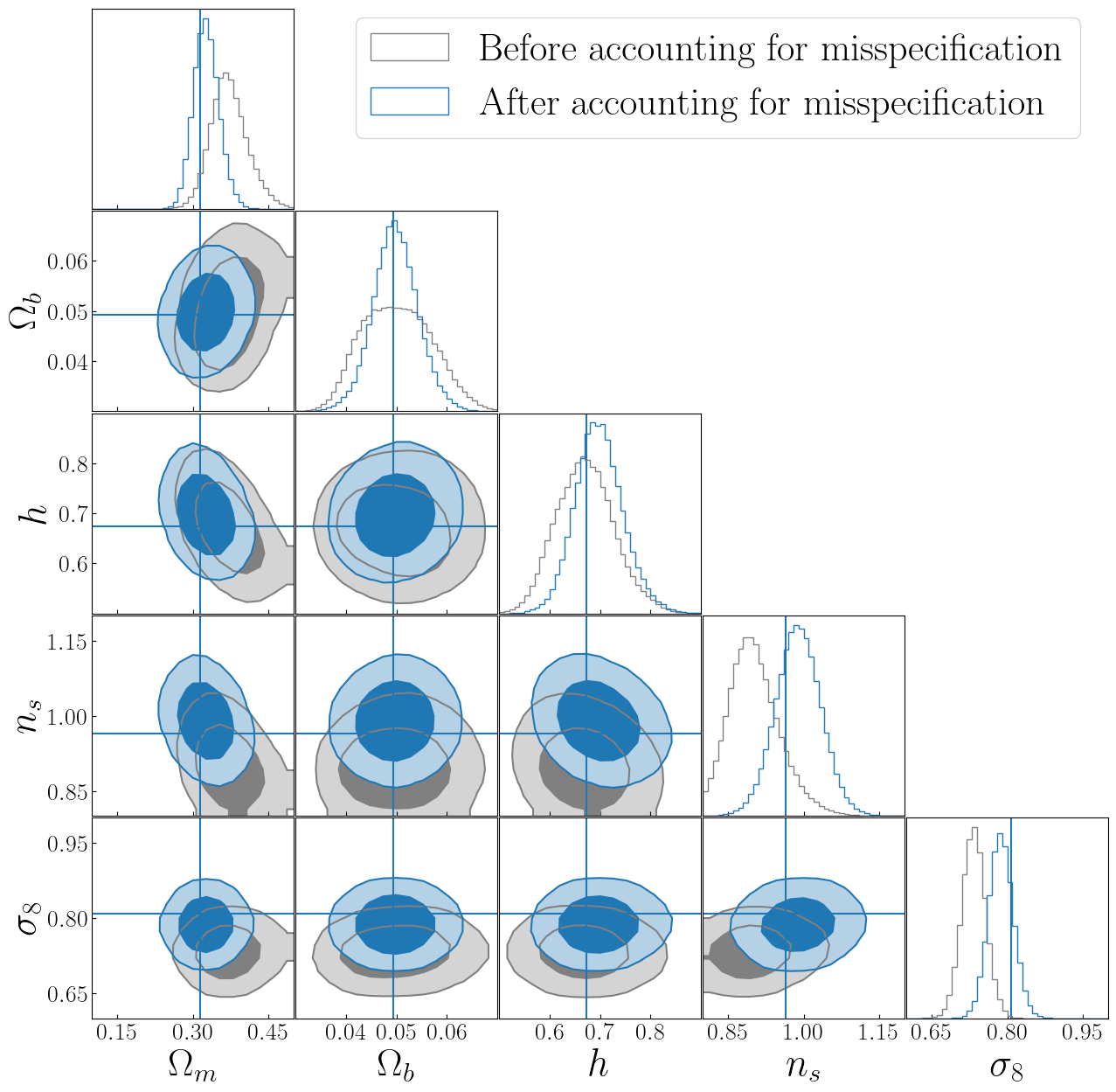}
\caption{Posterior distributions for the cosmological parameters $\Omega_m$, $\Omega_b$, $h$, $n_s$, and $\sigma_8$ inferred from a single \texttt{TEST2} simulation. Contours indicate the 68th and 95th percentiles. The true parameter values used to generate the simulation are marked by the intersection of the blue lines. Blue and grey contours show the posteriors before and after applying our method, respectively. The blue posterior appears more compatible with the true cosmological parameters, highlighting the positive impact of our method.}
\label{fig:posterior_comparison_validation}
\end{figure}

We train our pipeline on this setup and compare the resulting posteriors before and after applying our method.\footnote{The optimal regularization strength $\lambda$ in this setting is found to be $\lambda = 1$.} The comparison is shown in Fig.~\ref{fig:posterior_comparison_validation}. The updated inference appears more consistent with the fiducial value, particularly for the $\sigma_{8}$ parameter.

\subsection{Inference on BOSS data}
\label{sec:obs_results}

\subsubsection{Cosmological Constraints}

\begin{figure}
\includegraphics[scale=0.28]{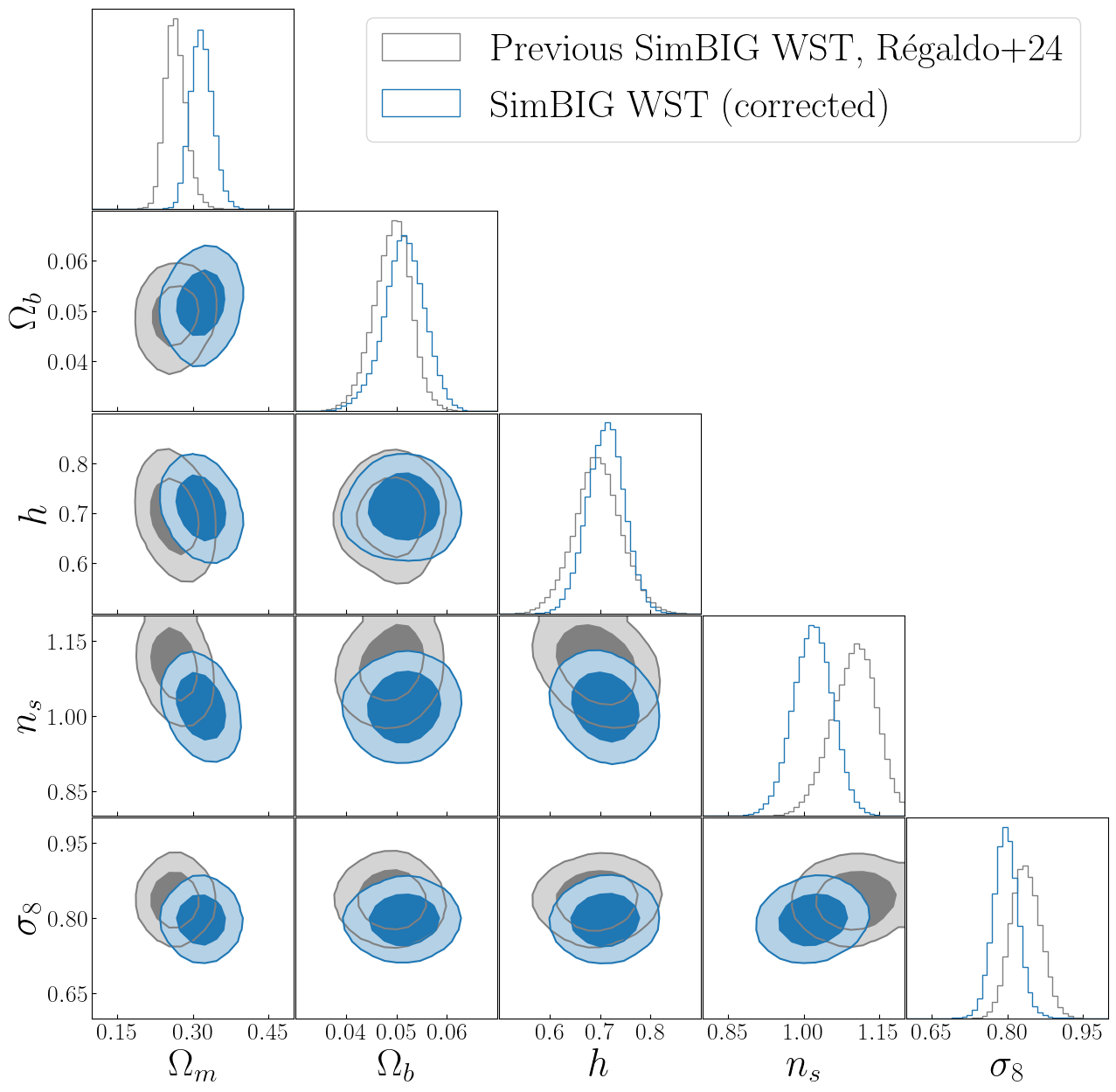}
\caption{Same as Fig.~\ref{fig:posterior_comparison_validation} but for the BOSS data.}
\label{fig:Difference of inference after regularization}
\end{figure}

We now present the results obtained when accounting for model misspecification in the analysis of the BOSS galaxy sample. In this setting, the optimal regularization parameter is found to be $\lambda = 1$. Fig~\ref{fig:Difference of inference after regularization} shows the resulting posteriors before and after applying our method. The blue contours correspond to our updated inference, while the grey contours reproduce the results from \cite{SimBIG.WST}. We observe substantial shifts in the posteriors, particularly for $\Omega_{m}$, $n_{s}$, and $\sigma_{8}$. Additionally, despite discarding some of the summary statistic coefficients, our posterior is slightly tighter than the original \simbig~WST analysis. This improvement, also observed in Fig.~\ref{fig:posterior_comparison_validation}, stems from a reduction in variability across the ensemble of NPEs.
We illustrate this in Fig.~\ref{fig:variability sigma8 OmegaM} where the left panel presents three of the NPEs included in the original ensemble and the right panel presents the NPEs after applying our model misspecification correction. 
While the individual posteriors of the NPEs are comparable, they are more consistent and thus yields a more constrained ensemble.

\begin{figure}
\includegraphics[scale=0.28]{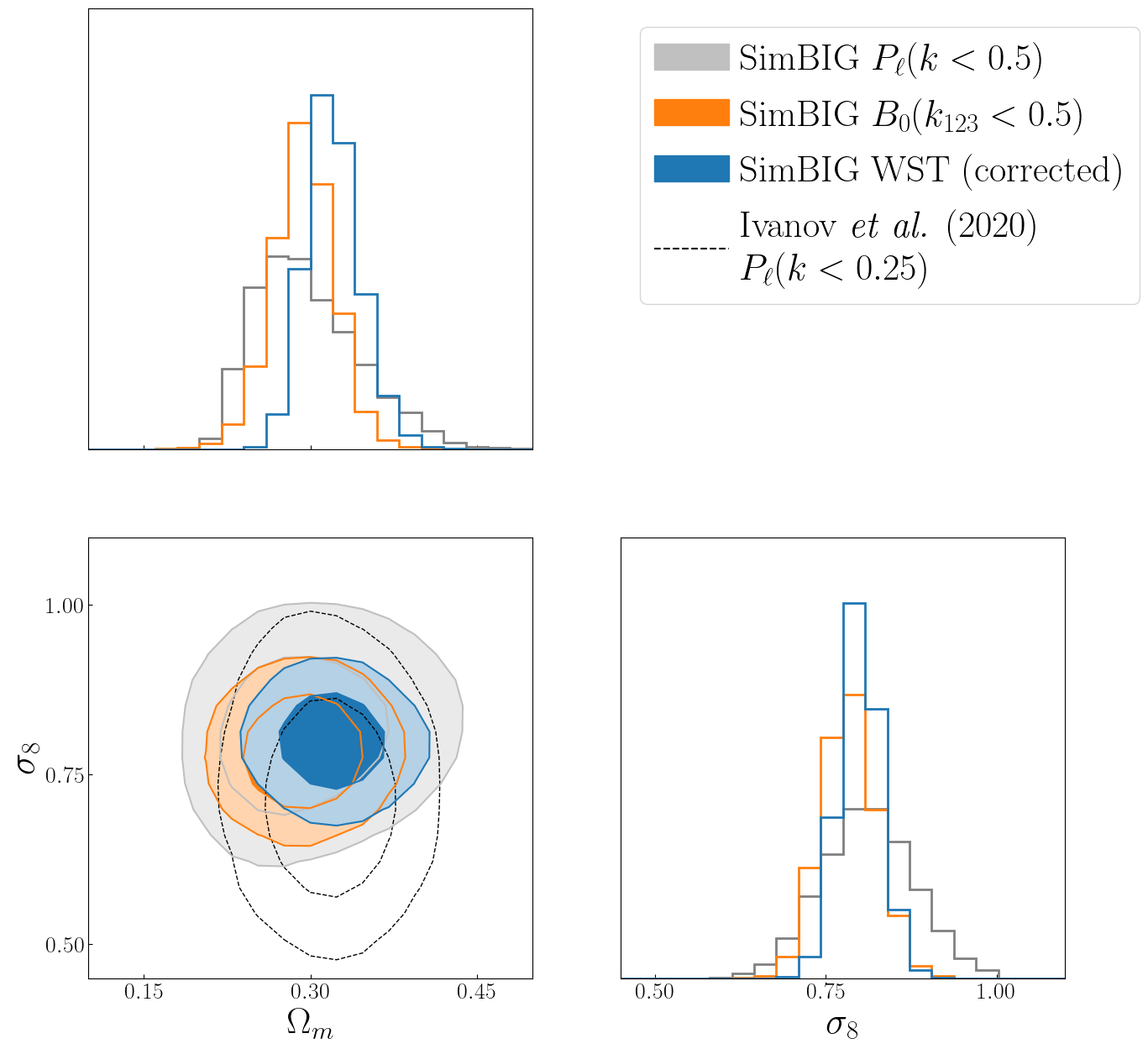}
\caption{
Updated cosmological constraints on $\Omega_m$ and $\sigma_8$ for the \simbig~WST analysis using our method to correct for model misspecification (blue). 
We find that $\lambda = 1$ is the optimal value for our correction. 
The contours represent the 68th and 95th percentile of the posteriors.
For comparison, we include constraints from: 
the \simbig~power spectrum~\citep[grey;][]{Hahn2023_pnas}, 
the \simbig~bispectrum~\citep[orange;][]{Hahn2024}, and the standard power spectrum analysis using perturbation theory~\citep[dotted;][]{Ivanov2020}. 
The corrected \simbig~WST analysis provide significantly more precise, yet consistent, constraints on both $\Omega_m$ and $\sigma_8$. 
}
\label{fig:final result BOSS}
\end{figure}

We present our cosmological constraints on the $\Lambda$CDM parameters in Table~\ref{table:obs_stats} with and without priors from Big Bang Nucleosynthesis (BBN). 
Furthermore, in Fig.~\ref{fig:final result BOSS}, we compare our constraints on $\Omega_m$ and $\sigma_8$ (blue) to: the \simbig~power spectrum analysis~\citep[grey;][]{Hahn2023_pnas}, 
the \simbig~bispectrum analysis~\citep[orange;][]{Hahn2024}, and the standard power spectrum analysis using perturbation theory~\citep[dotted;][]{Ivanov2020}. 
The contours indicate the 68th and 95th percentile of the posteriors. We also show in Fig.~\ref{fig:full plot final result} the full set of constraints, including $h$, $n_s$, and $\Omega_b$.

Overall, our constraints are consistent and significantly more precise than the other analyses. 
Our $\Omega_m$ and $\sigma_8$ constraints are  $1.4$ and $3.1$ times more precise than the standard power spectrum analysis using perturbation theory. 
In fact, our constraints are even more precise than the \simbig~bispectrum analysis.
This suggests that the WST contains an amount of additional cosmological information beyond the 2 and 3-point functions, consistent with previous forecasts~\cite{Eickenberg2022, Valogiannis2022a}.

From our $\Lambda$CDM posterior, we can derive constraints on the Hubble constant, $H_0$, and $S_8 = \sigma_8 \sqrt{\Omega_m / 0.3}$ to inform recent cosmological tensions~\cite[e.g.,][]{Abdalla2022}. 
Combining our posterior with a prior from BBN~\citep{Aver2015, Cooke2018, Schoneberg2019}, we get $H_0 = 68.8^{+2.8}_{-2.6}$ and $S_8=0.83^{+0.04}_{-0.04}$ . 
We find a lower value of $H_0$ that is consistent with other galaxy clustering analyses and in good agreement with experiments using the cosmic microwave background (CMB). 
Meanwhile, we find a slightly higher value of $S_8$ compared to recent galaxy clustering analyses~\citep[e.g.,][]{philcox2022, chen2022}.
This is in good agreement with CMB analyses, but in slight tension with some weak lensing experiments~\citep[e.g.,][]{asgari2021, amon2022, secco2022}. A comparison between our results and other analyses is available in Fig.~\ref{fig:comparison S8 H0}.

\begin{table*}
    \def\arraystretch{1.5}
    \centering
    \begin{tabular}{c|cccccc}
        \hline
        \hline
        Analysis & $\Omega_m$ & $\Omega_b$ & $h$ & $n_s$ & $\sigma_8$ \\
        \hline
        \hline
        Corrected {\sc SimBIG}~WST (uniform prior)
         & $0.32_{-0.02}^{+0.02}$ & $0.051_{-0.004}^{+0.004}$ & $0.71_{-0.04}^{+0.04}$ & $1.02_{-0.04}^{+0.04}$ & $0.80_{-0.02}^{+0.02}$ \\ 
        \hline
        Corrected {\sc SimBIG}~WST (BBN prior)
 & $0.32_{-0.02}^{+0.02}$ & $0.049_{-0.003}^{+0.004}$ & $0.68_{-0.02}^{+0.02}$ & $1.02_{-0.04}^{+0.04}$ & $0.79_{-0.02}^{+0.03}$ \\
    \hline
         \hline
    \end{tabular}
    \caption{Constraints on the $\Lambda$CDM cosmological parameters inferred from the BOSS data with our method, both with and without a BBN prior. For each parameter, we show the 68\% credible interval centered on the median.}
    \label{table:obs_stats}
\end{table*}

\subsubsection{Robustness to Choice of Prior}
\label{subsub:Dependence on the prior}

Finally, we assess whether our method is sensitive to the choice of prior distribution. A legitimate concern is that the method could bias the observational posterior toward the center of the prior. Specifically, if the distance between the observation's statistics and those of the training set is minimized too aggressively, the learned transformation might push the observation toward the centroid of the training distribution. In such a scenario, although the variability in the posteriors would be reduced --- allowing us to trivially pass the test in Sect.~\ref{subsub: Variability of posteriors} --- the inference would be obviously biased.

\begin{figure}
\includegraphics[scale=0.28]{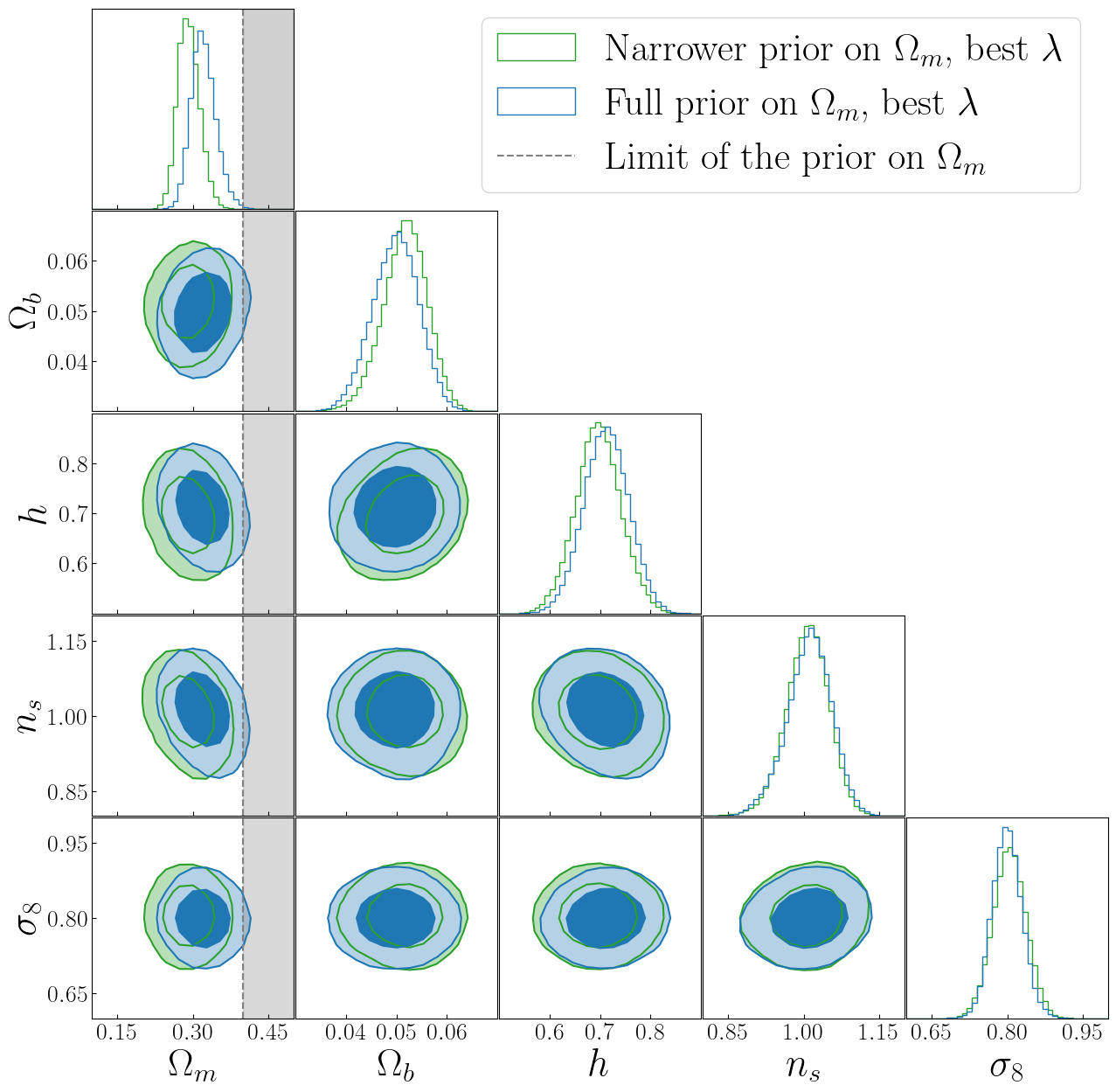}
\caption{Posterior distributions for the cosmological parameters $\Omega_m$, $\Omega_b$, $h$, $n_s$, and $\sigma_8$ inferred from the BOSS galaxy sample. Contours indicate the 68th and 95th percentiles. The blue posterior is obtained using a training set with a narrower uniform prior on $\Omega_m$ compared to the green posterior. The consistency between the two posteriors indicates that our method is robust to changes in the prior and does not collapse toward its center.}
\label{fig:Comparison prior cut or not} 
\end{figure}

To test this, we consider an alternative training set using the same forward model and the same number of simulations, but with a narrower uniform prior on $\Omega_{m}$: $[0.1, 0.4]$ instead of $[0.1, 0.5]$. We then repeat the training procedure and inference, and show the resulting observational posterior in Fig.~\ref{fig:Comparison prior cut or not}. We observe that the inferred posterior for $\Omega_{m}$ remains consistent between the two priors. In particular, it does not collapse toward the center of the narrower prior (0.25), confirming that our method does not induce artificial bias from the prior distribution. Empirically, we do observe that setting $\lambda$ too high can lead to such a collapse. This highlights the importance of a careful selection procedure for $\lambda$.

\section{Conclusion}
\label{sec:conclusion}

In this work, we investigated and addressed model misspecification challenges arising in SBI analyses of galaxy clustering data. 
We focused on a specific use case, analyzing a BOSS galaxy sample with the {\sc SimBIG} SBI framework using the WST summary statistics. Following up on a past analysis~\cite{SimBIG.WST} that revealed model misspecification issues in this context, we made the following contributions:
\begin{itemize}
    \item We focused on settings where model misspecification can be reliably detected from a single observational sample, i.e., where the observational data appears OOD with respect to the simulated data. In this context, we emphasized the crucial role of the summary statistics, which may capture unrealistic simulator-specific features, leading to unreliable predictions. We highlighted the importance of systematically quantifying the sensitivity of these summary statistics to plausible simulator variations. 
    Additionally, we introduced a practical diagnostic to detect OOD behavior and quantify its impact on inference, based on a measure of the variability of observational posteriors obtained from equally well trained NPEs.
    \item We introduced a two-step method to address model misspecification in this context. The first step removes summary statistic components found to be most sensitive to plausible simulator variations. The second step 
    learns a transformation of the pruned statistics, reducing the MMD between the observational and training data, while retaining as much as possible information on the target cosmological parameters. A critical aspect of this step is the careful balancing between robustness and informativeness, controlled by a scalar hyperparameter, $\lambda$. We discussed --- and cautioned --- how strongly the chosen value of $\lambda$ influences the learned statistics and thus inference outcomes, and provided a systematic method for optimally tuning this hyperparameter.
    \item We applied our method to analyze the BOSS sample with \simbig~and WST statistics and resolved the previously identified misspecification issues~\cite{SimBIG.WST}.  We presented improved cosmological parameter constraints. With uniform priors on the cosmological parameters, our analysis yielded constraints of $\Omega_m = 0.32^{+0.02}_{-0.02}$ and $\sigma_8 = 0.80^{+0.02}_{-0.02}$ (median and 68\% credible intervals). These constraints are consistent with other galaxy clustering analyses and are respectively $1.4$ and $3.1$ times tighter than those obtained from a standard perturbation-theory-based power spectrum analysis~\cite{Ivanov2020}. They also appear slightly more precise than an equivalent {\sc SimBIG} bispectrum analysis~\cite{Hahn2024}. This confirms the significant additional cosmological information encoded in the WST, in line with the improvements anticipated from prior Fisher forecasts~\cite{Eickenberg2022, Valogiannis2022a}.
\end{itemize}

While our method effectively addresses model misspecification, selecting the regularization strength $\lambda$ remains a practical challenge. The current approach requires training multiple neural density estimators across a range of $\lambda$ values, which can be computationally expensive. This limitation could be alleviated in future work with more efficient strategies for $\lambda$ selection.

More critically, our method currently sacrifices interpretability: by learning a highly flexible transformation of the summary statistics, we lose the ability to relate the inference to physically meaningful features of the data. In the case of WST, this means we can no longer interpret the cosmological constraints in terms of a multiscale representation of the galaxy clustering signal. While this trade-off enables more robust inference under misspecification, it can limit the insight we can extract from the data and makes it harder to identify which features of the simulations may require improvement. Addressing this limitation is a key direction for future work. In particular, we plan to investigate restricted classes of transformations that preserve interpretability — for instance, by constraining $\eta$ to act locally or hierarchically on multiscale coefficients, retaining a correspondence between transformed features and physical scales.

We also note that our method breaks the amortization of the inference, as the summary statistics transformation becomes tailored to the specific observation. However, in settings with multiple observations, one could use a subset of them to learn the transformation, and then apply it to the remaining observations, thereby restoring amortized inference.

Although our work focuses on SBI, similar model misspecification challenges will also arise in emulator-based approaches involving higher-order statistics~\cite[e.g.,][]{Valogiannis2024}, which similarly rely on neural networks trained on synthetic data. Extending the kinds of diagnostics and mitigation strategies explored here to that context could be a promising direction for future work.

Finally, while our approach offers a practical solution to mitigate model misspecification, it should not be seen as a substitute for improving the forward model itself. Instead, robust cosmological SBI should rely on both continued improvements in the simulator fidelity and diagnostic tools and correction methods like ours. Together, these efforts can ensure that cosmological inference remains both accurate and resilient to modeling uncertainties.

\begin{acknowledgments}
We acknowledge fruitful discussions with Chirag Modi, Yuling Yao, Alberto Bietti, and François Lanusse, and thank Adrian Bayer for helpful comments on the manuscript. We also gratefully acknowledge the Flatiron Institute and the Scientific Computing Core for their support.
\end{acknowledgments}

\bibliography{bib}

\appendix

\onecolumngrid

\newpage
\section{Additional Figures}
\label{app:add_figs}

\begin{figure*}[h]
\centering
\begin{tikzpicture}[
    layer/.style={rectangle, draw=black, minimum width=2.5cm, minimum height=1.2cm, align=center},
    arrow/.style={->, thick},
    node distance=1.0cm and 0.5cm
]

\node[layer] (input) {$s$ \\ (2073)};
\node[layer, right=of input] (fc1) {Linear Layer \\ (2073 → 2073)};
\node[layer, right=of fc1] (bn1) {BatchNorm};
\node[layer, right=of bn1] (relu1) {ReLU};
\node[layer, right=of relu1] (fc2) {Linear Layer \\ (2073 → 2073)};
\node[layer, right=of fc2] (output) {$\eta_{\psi}$(s) \\ (2073)};

\draw[arrow] (input) -- (fc1);
\draw[arrow] (fc1) -- (bn1);
\draw[arrow] (bn1) -- (relu1);
\draw[arrow] (relu1) -- (fc2);
\draw[arrow] (fc2) -- (output);

\end{tikzpicture}
\caption{Schematic of the $\eta_{\psi}$ network, which transforms the summary statistics $s$ (of dimension 2073) into $\eta_{\psi}(s)$. The architecture consists of two linear layers that preserve the input dimension, with a BatchNorm layer and ReLU activation applied after the first layer.}
\label{fig:architecture}
\end{figure*}
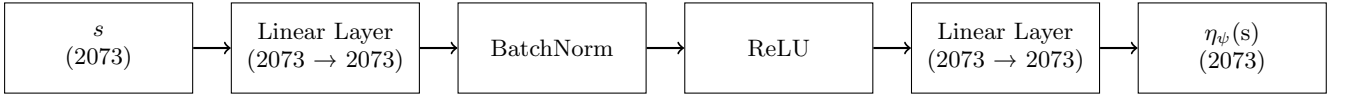

\begin{figure*}[h]
\includegraphics[scale=0.5]{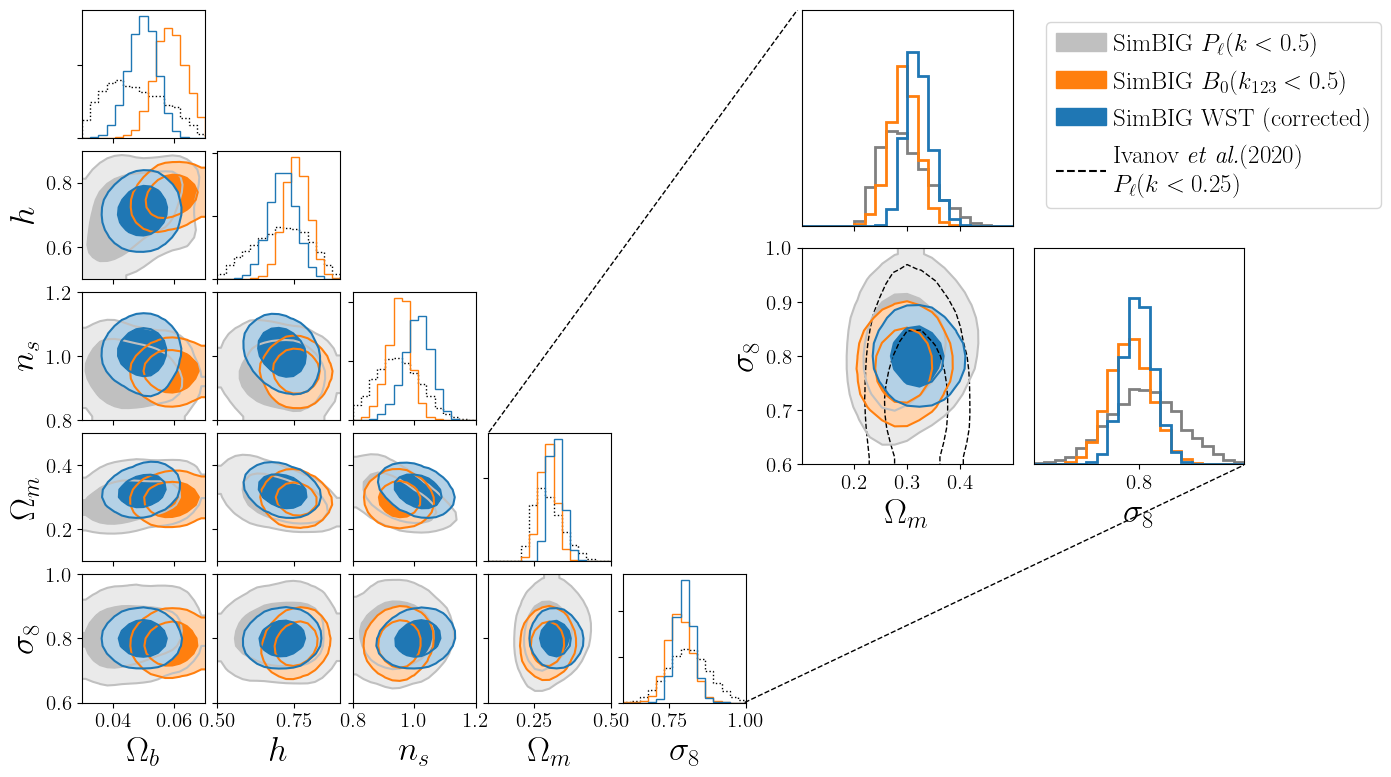}
\caption{Same as Fig.~\ref{fig:final result BOSS} but for the five cosmological parameters $\Omega_m$, $\Omega_b$, $h$, $n_s$, and $\sigma_8$.}
\label{fig:full plot final result}
\end{figure*}

\begin{figure*}
\includegraphics[scale=0.6]{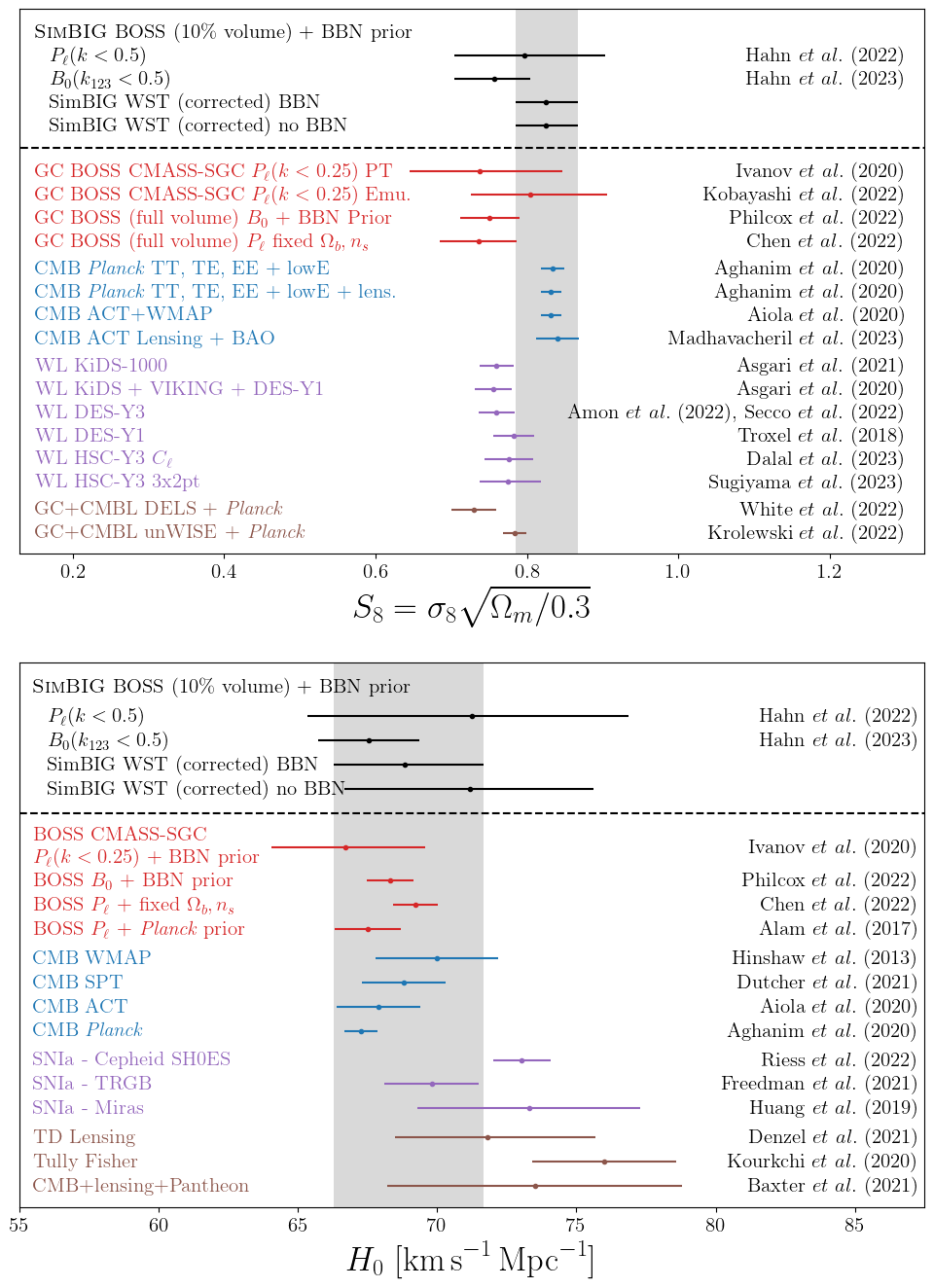}
\caption{Comparison of the \simbig ~$S_8$ (top) and $H_0$ (bottom) constraints (black) with existing bounds in the literature. We include constraints from galaxy clustering (red), CMB (blue), weak lensing (WL, purple), clusters (pink) and multiprobe (brown) analyses.}
\label{fig:comparison S8 H0}
\end{figure*}

\end{document}